\documentclass[11pt]{revtex4-1}

\usepackage{graphicx}
\usepackage{natbib}
\usepackage{amsmath}

\begin{document}
\title{Consequences of fluctuating group size for the evolution of cooperation}


\author{{\AA}ke Br\"{a}nnstr\"{o}m}
\affiliation{Department of Mathematics and Mathematical Statistics, Ume\aa{} University, SE-90187, Ume\aa{}, Sweden}

\author{Thilo Gross}
\affiliation{Max-Planck-Institute for Physics of Complex Systems, N\"othnitzer Stra\ss{}e 38, 01187 Dresden, Germany}

\author{Bernd Blasius}
\affiliation{Institute for Chemistry and Biology of Marine Environment, Oldenburg University, 26111 Oldenburg, Germany}

\author{Ulf Dieckmann}
\affiliation{Evolution and Ecology Program, International Institute for Applied Systems Analysis (IIASA), Schlossplatz~1, 2361~Laxenburg, Austria}

\date{\today}

\begin{abstract}
{ \small
Studies of cooperation have traditionally focused on discrete games such as 
the well-known prisoner's dilemma, in which players choose between two pure 
strategies: cooperation and defection. Increasingly, however, cooperation is 
being studied in continuous games that feature a continuum of strategies 
determining the level of cooperative investment. For the continuous 
snowdrift game, it has been shown that a gradually evolving monomorphic 
population may undergo evolutionary branching, resulting in the emergence of 
a defector strategy that coexists with a cooperator strategy. This 
phenomenon has been dubbed the `tragedy of the commune'. Here we study the 
effects of fluctuating group size on the tragedy of the commune and derive 
analytical conditions for evolutionary branching. Our results show that the 
effects of fluctuating group size on evolutionary dynamics critically depend 
on the structure of payoff functions. For games with additively separable 
benefits and costs, fluctuations in group size make evolutionary branching 
less likely, and sufficiently large fluctuations in group size can always 
turn an evolutionary branching point into a locally evolutionarily stable 
strategy. For games with multiplicatively separable benefits and costs, 
fluctuations in group size can either prevent or induce the tragedy of the 
commune. For games with general interactions between benefits and costs, we 
derive a general classification scheme based on second derivatives of the 
payoff function, to elucidate when fluctuations in group size help or hinder 
cooperation.}
\end{abstract}

\maketitle

\section{Introduction}
Cooperation is ubiquitous in nature, and the cooperative integration
of lower-level entities into higher-level units has been vital for the
development of life on earth \citep{32}. While cooperation in the
broad sense only implies joint action, the term is often used more
strictly to describe situations in which cooperators help others at a
cost to themselves. These interactions are typically vulnerable to
cheating and exploitation by defectors that benefit without making
costly cooperative contributions of their own. Cheating and
exploitation are observed in viruses \citep{46}, bacteria \citep{38},
yeast \citep{17}, amoebas \citep{6, 7, 42}, fish \citep{37}, and humans
\citep{1}.

How cooperation can persist in the presence of cheaters is not obvious. At 
first glance it often appears as though the fitness of cheaters exceeds that 
of cooperators. Indeed, the well-known tragedy of the commons \citep{21} shows 
that even when cooperation is beneficial for a group, selection acting at 
the individual level often eliminates cooperation altogether. This has 
attracted significant scientific interest throughout the last decades, with 
explanations proposed for the origin and maintenance of cooperation falling 
into three main categories. First, kin selection \citep{18, 19, 20} successfully 
explains many forms of cooperation among genetically related individuals. 
Second, selection at the level of groups \citep{47, 48}, through which 
subpopulations with non-cooperative individuals are at a reproductive 
disadvantage, promotes cooperation under certain conditions. Third, direct 
and indirect reciprocation have been shown to foster cooperation \citep{45, 2, 1}. 
These alternative mechanisms are further discussed in \citet{35}.

Most game-theoretical studies of cooperation fall into the third
category described above and revolve around a game known as the
prisoner's dilemma \citep{2}. The classic variant of this game is
played by two players choosing between two pure strategies,
cooperation or defection, but the game can be generalized to an
arbitrary number of players \citep{27, 9} and to continuous levels of
cooperative contributions \citep{30, 28}. In the latter case,
cooperative investments vary continuously and are represented by real
numbers, denoted here by $r_1 $ or $r_2 $. The payoff of an $r_1
$-strategist facing an $r_2 $-strategist is $B(r_2 )-C(r_1 )$, where
$B$ and $C$ areq increasing functions that quantify the
benefits and costs of cooperative investments. Since cooperative
investments do not directly benefit the acting individual, defection
is the rational choice when the game is played only once. In many
cases, however, it is more reasonable to assume that all players
benefit equally from cooperative investments. For example, the
digestive enzymes produced by a cell of the yeast
\textit{Saccharomyces cerevisiae} can be used by all nearby cells,
including the producing cell itself \citep{16}.  Likewise, while the stalk
produced by the amoeba \textit{Dictyostelium discoideum} can be
exploited by cheaters, it also vitally benefits the
cooperators. Further examples of processes resulting in shared
benefits are cooperative hunting, vigilance behavior, group foraging,
and parental care \citep{29}. Situations in which individuals directly
benefit from cooperative acts that they perform can be described by
the snowdrift game \citep{43}, synonymously known as the hawk-dove game \citep{31}
or chicken game \citep{40}.

To better understand the evolution of cooperation and defection when
all players are benefiting from cooperative investments, \citet{10}
studied the snowdrift game with continuous investments. In this game,
the payoff of an $r_1 $-strategist facing an $r_2 $-strategist is
$B(r_1 +r_2 )-C(r_1 )$, where the functions $B$ and $C$ are chosen so
that cooperation is more successful than defection in groups of
defectors, but defection is advantageous in groups of
cooperators. Consequently, cooperation in the snowdrift game always
develops to some intermediate degree. However, assuming small
mutations in continuous cooperative investments, \citet{10} showed
that this gradual buildup of cooperation was sometimes followed by the
emergence of cheaters with little or no cooperative investments, while
the remaining cooperators became even more cooperative. Similar
results have been obtained by \citet{3} in the context of the social
amoeba \textit{Dictyostelium}. When starvation is imminent, one or
several strains of this amoeba aggregate to form fruiting bodies that
enable spore dispersal \citep{39, 11}. A strain, however, may forego
investing into the stalk of the fruiting body and instead take a free
ride on the investments of other strains. \citet{3} modeled this
process with spores as players and with the investment of strains into the
construction of stalks as strategies. In this model, the payoff of an
$r_1 $-strategist facing an $r_2 $-strategist is multiplicative,
$B(r_1 +r_2 )C(r_1 )$, with $B$ an increasing and $C$ a decreasing
function of cooperative investments. They showed that fluctuation in
player numbers resulted in evolutionary branching and in the
subsequent emergence and coexistence of low-investing cheaters and
high-investing cooperators.

\begin{figure}
\centering
\begin{tabular}{ll}
\raisebox{5cm}{(a)}
\includegraphics[width=0.3\textwidth]{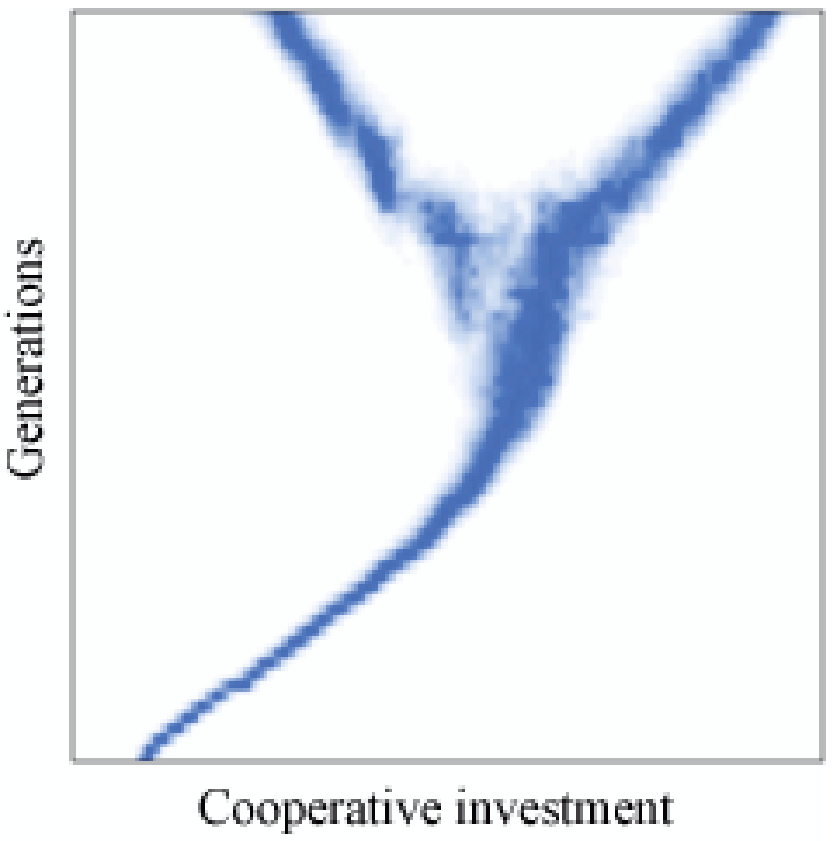} &
\raisebox{5cm}{(b)}
\includegraphics[width=0.3\textwidth]{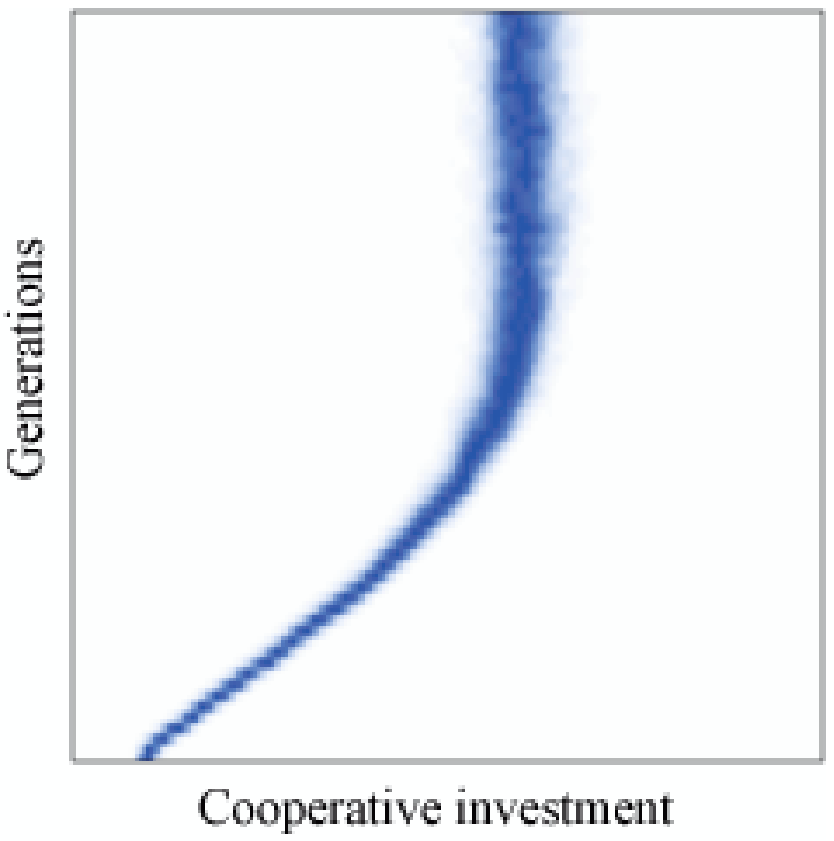}
\end{tabular}
\caption{Individual-based simulations of a multi-player version of the
  continuous investment game proposed by \citet{10}. Parameters used here
  are identical to those used in Fig. 1A of \citet{10}. (a) Game played as a
  two-player game: evolutionary branching leads to the emergence of
  cheaters, a characteristic process \citet{10} dubbed the tragedy of the
  commune. (b) Game played with a random number of players (with
  interactions taking place among either 1 or 3 players with equal
  probability): although the average number of players is the same as
  before, evolutionary branching does not occur, and hence the tragedy
  of the commune is avoided. Note that although games with 1 player
  may seem odd, a natural interpretation of such games often exists
  for specific examples, e.g. single-clone aggregation in the case of
  \emph{Dictyostelium}.}
\label{fig1}
\end{figure}

The work of \citet{10} and \citet{3} shows that selection on levels of
cooperative investments need not always be stabilizing. Rather,
initially monomorphic populations evolving in cooperation games may
experience disruptive selection, resulting in evolutionary branching
and the emergence of dimorphic evolutionary outcomes in which
low-investing and high-investing individuals coexist. \citet{10}
investigated this evolutionary phenomenon, which they dubbed the
\textit{tragedy of the commune}, for games played in two-player
groups. However, in many situations it is more natural to expect that
interactions take place in groups of fluctuating size, for example, as
a consequence of abstaining or of local interactions coupled with
dispersal or movement. Since environmental fluctuations have been
shown to promote the coexistence of competing populations \citep{26} and to
facilitate evolutionary branching in some models, such as the
site-based model studied by \citet{14}, one might expect that fluctuating
group size would render the tragedy of the commune more likely.  This,
however, need not be the case. Figure~\ref{fig1} shows a multi-player extension
of a game considered by \citet{10}, in which players interact in randomly
formed groups that change between each interaction. When the size of
these groups changes significantly from one interaction to the next,
the tragedy of the commune no longer occurs.

The aim of this paper is to explore the evolutionary consequence of 
fluctuating group size for cooperation games. We first define a large class 
of games that includes the snowdrift game considered by \citet{10}, the 
\textit{Dictyostelium} model conceived by \citet{3}, the prisoner's dilemma, the stag-hunt game, and 
other public-good (joint-effort) games. For this class of games, we explore 
the evolutionary consequences of fluctuating group size for the 
establishment of cooperation and the tragedy of the commune.

\section{Cooperation games with fluctuating group size}
In this section we first explain why fluctuations in the size of groups of 
interacting players are generically expected in nature. We then introduce a 
class of cooperation games with continuous investments that incorporate 
fluctuating group size. From the demographic dynamics resulting from games 
in this class we determine the initial growth rate of a rare mutant 
strategy. This lays the foundation for our analysis of the evolutionary 
dynamics of cooperative investments. Finally, we describe the potential 
outcomes of gradual evolution in a monomorphic population with at most one 
interior evolutionarily singular strategy.

\subsection{Fluctuating group size}

\begin{figure}
\centering
\begin{tabular}{ll}
\raisebox{4.5cm}{(a)}
\includegraphics[width=0.4\textwidth]{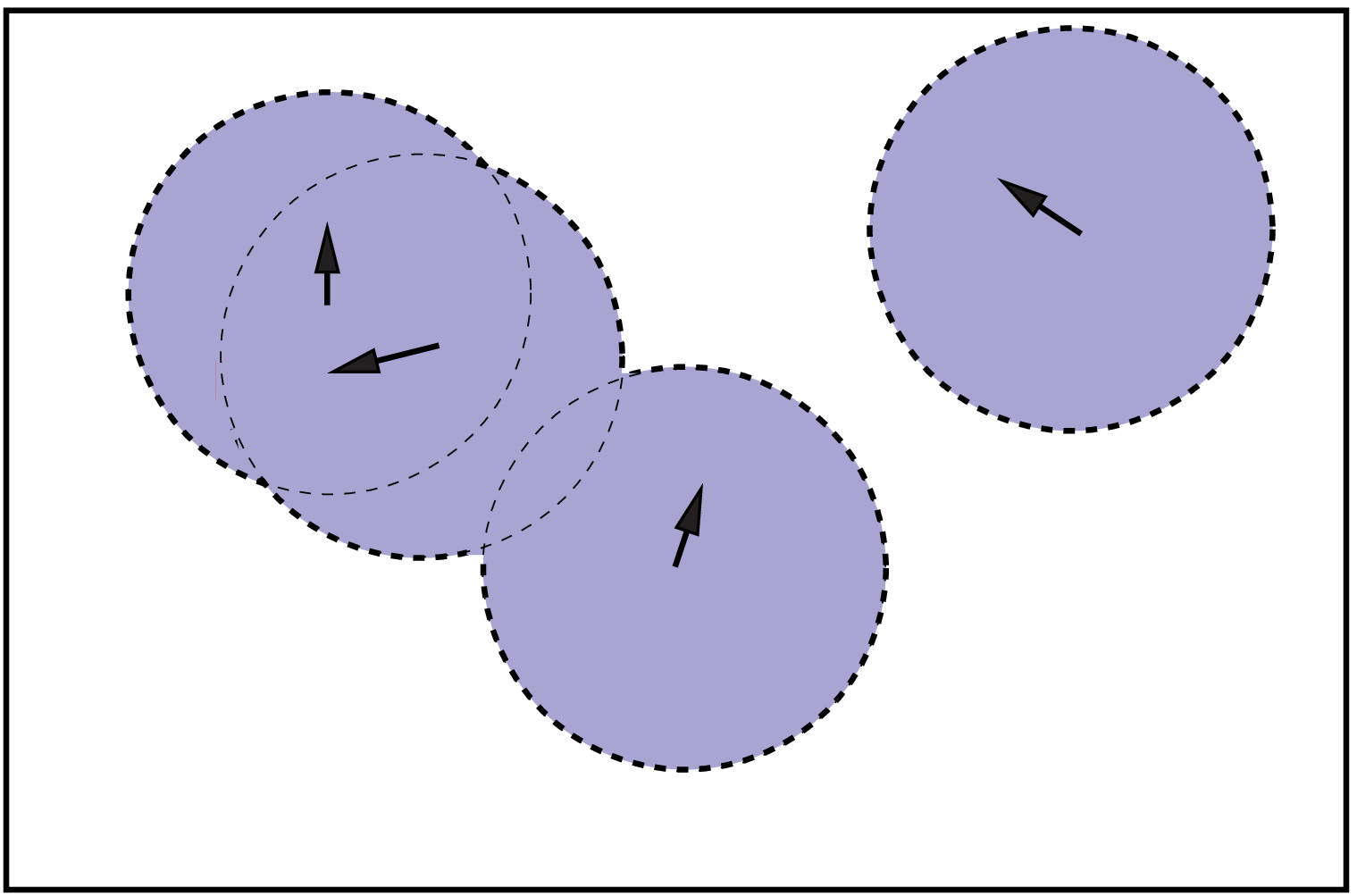} &
\raisebox{4.5cm}{(b)}
\includegraphics[width=0.4\textwidth]{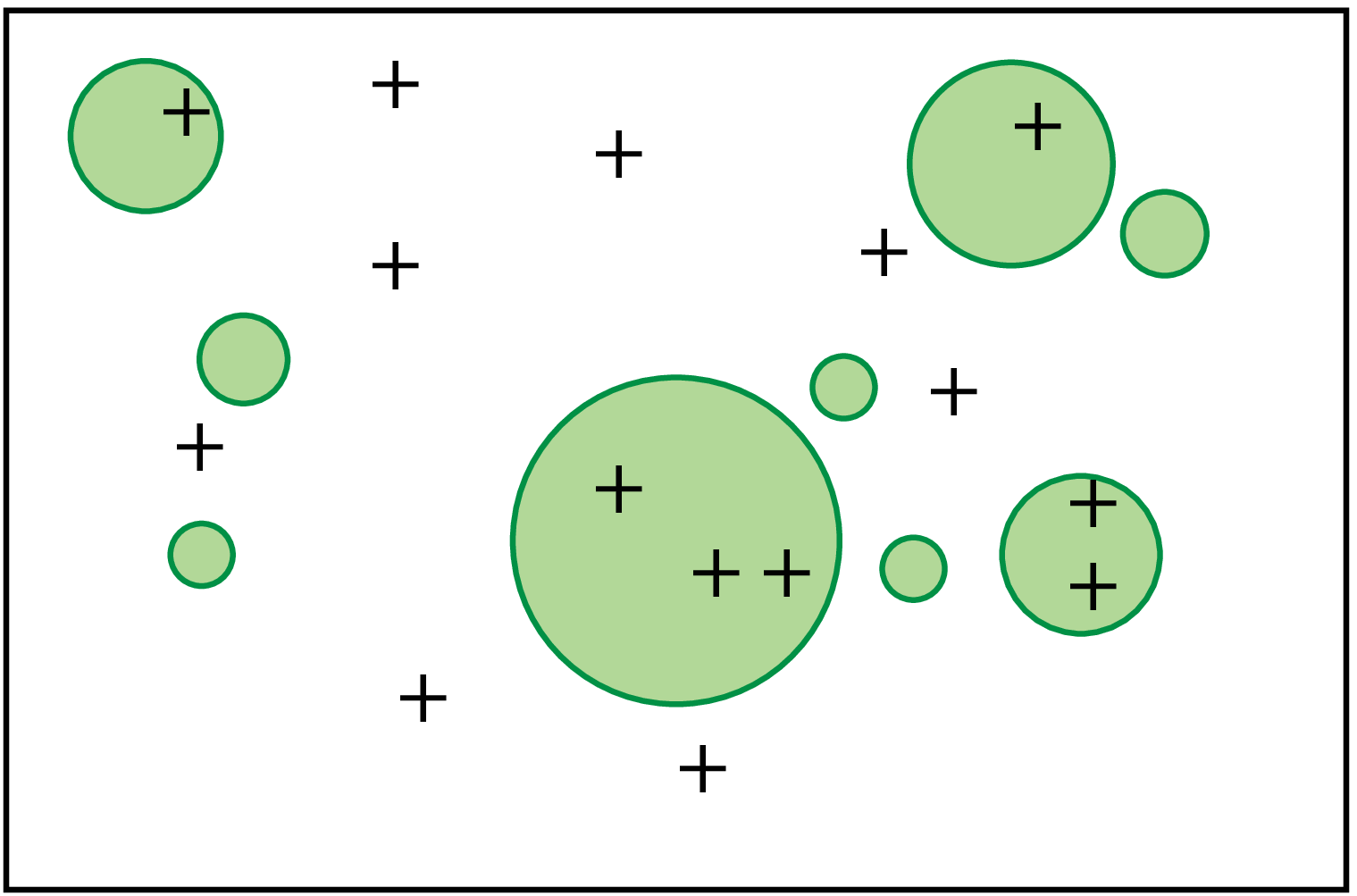} 
\end{tabular}
\caption{Two examples of processes giving rise to fluctuations in the
  size of groups of interacting players. (a) Movement of individuals
  in conjunction with infrequent local interactions between nearby
  individuals give rise to games in which groups are formed through
  the overlap of interaction zones. (b) Players are distributed over
  an area, with interactions occurring among those players that occupy
  the highlighted patches. A specific example is the dispersal of
  spores or seeds over an area containing many disconnected patches of
  suitable habitat.}
\label{fig2}
\end{figure}

Figure~\ref{fig2} depicts two situations in which variation in group size naturally 
occurs. First, movement and infrequent interactions among nearby players 
leads to a class of games in which groups are formed through the overlap of 
interaction zones. Although Fig.~\ref{fig2}a is most easily interpreted in terms of 
binary interactions, the interaction strengths among players in general 
depend on factors such as their distance. Second, Fig.~\ref{fig2}b shows how games 
are formed when players are repeatedly distributed onto patches of different 
sizes, giving rise to distinct groups of interacting players. Significant 
variation in patch size leads to a wide distribution of the number of 
players in a group. This situation arises, for example, when spores or seeds 
are dispersed over an area with fragmented patches of suitable habitat and 
subsequently interact within each patch. Our analysis below encompasses both 
situations depicted in Fig.~\ref{fig2}.

\subsection{Payoffs in cooperation games}
When a group of $k$ players has been formed, we assume that each player 
contributes an amount or effort $r_i$ towards the group's total effort $r_1 
+\ldots +r_k$. The contribution $r_i$ is the strategy or trait value of a 
player and may optionally be constrained to an interval, e.g. $0\le r_i \le 
1$. We assume that the payoff $P(r_i ,r_{\rm{s}} ,k)$ for a focal individual 
playing strategy $r_i$ in a group of players with strategies $r_1 ,\ldots 
,r_k$ may depend on the focal player's own contribution $r_i$, on the 
focal player's share of the total effort, $r_{\rm{s}} =(r_1 +\ldots +r_k 
)/k$, and on the number of individuals $k$ in the group. By choosing $P$ 
appropriately, we recover all traditionally studied cooperation games as 
special cases. For example, the two-player prisoner's dilemma is obtained by 
choosing $P(r_i ,r_{\rm{s}} ,k)=B(kr_{\rm{s}} -r_i )-C(r_i )$ with 
increasing functions $B$ and $C$, and the $k$-player public-good 
(joint-effort) game by choosing $P(r_i ,r_{\rm{s}} ,k)=mr_{\rm{s}} -r_i$ 
with a positive factor $m$.

We say that benefits and costs are additively or multiplicatively
separable if, respectively, $P(r_i ,r_{\rm{s}} ,k)=B(r_{\rm{s}}
,k)-C(r_i ,k)$, or $P(r_i ,r_{\rm{s}} ,k)=B(r_{\rm{s}} ,k)C(r_i
,k)$. For a fixed number of players, additively and multiplicatively
separable benefits and costs give rise to equivalent evolutionary
dynamics, as the multiplicatively separable payoff is additively
separable on a logarithmic scale. Importantly, however, the
evolutionary dynamics of these games is not equivalent in games with
fluctuating player numbers, as will become clear in Sect.~\ref{sec2.3} below.
The distinction between additive and multiplicative payoff structures
will play an important role in Sect.~\ref{sec3} when we examine the
effect of fluctuating group size on the tragedy of the commune. When
players pay a cost for making a cooperative contribution and benefit
from their group's total effort, it is natural to assume that the
payoff $P$ decreases with $r_i$ and increases with $r_{\rm{s}}$,
\begin{equation}
\label{eq1}
P_1 \left( {r_i ,r_{\rm{s}} ,k} \right)\le 0\;\mbox{and}\;P_2 \left( {r_i 
,r_{\rm{s}} ,k} \right)\ge 0,
\end{equation}
where the subscripts in $P_1$ and $P_2$ denote the partial derivatives of 
$P$ with respect to its first or second argument, respectively. Since this 
assumption is not needed for most of the arguments below, it will be invoked 
only when analyzing the sign of mixed derivatives of multiplicative payoff 
functions.

\subsection{Demographic dynamics in cooperation games} \label{sec2.3}
Based on the general specification of payoffs for players participating in 
cooperation games provided above, we now introduce the resultant demographic 
dynamics describing how player abundances change over time. For this we 
assume that, in successive generations, players are randomly distributed 
over groups of different size. The probability that an individual joins a 
game with $k$ participants is $p_k =kq_k /\left\langle k \right\rangle$, 
where $q_k $ is the fraction of groups with $k$ players and $\left\langle k 
\right\rangle =\sum_{k=1}^\infty k q_k$ is the average number of 
players in a group. Individuals then interact within the group and produce 
offspring in proportion to the payoff they received. Survival to the next 
generation is density-dependent, but independent of trait values. Under 
these assumptions, the per capita growth rate of an initially rare mutant 
strategy $m$ in an environment dominated by players with resident strategy 
$r$ is
\begin{equation}
\label{eq2}
f(r,m)=\sum_{k=1}^\infty p_k \left( P\left(m,\frac{m+(k-1)r}{k},k 
\right)-P(r,r,k) \right).
\end{equation}
In adaptive dynamics theory, this quantity is known as invasion
fitness \citep{33}. Equation (\ref{eq2}), the derivation of which is
provided in Appendix B, shows that for the cooperation games
considered here invasion fitness is given by the excess payoff of a
single mutant in groups of residents, relative to the payoff of a
resident in groups of residents, averaged according to the probability
$p_k$ that the mutant occurs in groups of size $k$. The determination
of this invasion fitness allows us to study the long-term evolution of
trait values under mutation and selection \citep{34, 8, 14} and thus forms
the foundation of our analysis of evolutionary dynamics in Sect.~\ref{sec3}.

Equation~(\ref{eq2}) shows that the payoffs players receive in groups
of different size are averaged arithmetically. Hence, if two or more
group sizes can occur, nonlinear transformations of the scale on which
payoffs are measured are not possible, since they would distort the
weighted average. Thus, only when the number of players is fixed, does
a transformation of payoffs to a logarithmic scale enable a reduction
from multiplicatively separable payoff structures to additively
separable payoff structures.

\subsection{Evolutionary dynamics in cooperation games}
When mutational steps are small and rare, resident communities will
successively be replaced by invading mutants with similar strategies
and positive invasion fitness. Driven by directional selection, this
process eventually ceases when evolution reaches either a boundary
strategy at which constraints prevent further evolution, or an
interior strategy at which selection pressures vanish. Strategies of
the latter type are called evolutionarily singular; in their vicinity,
an initially monomorphic population may experience disruptive
selection and thus become dimorphic. Figure~\ref{fig1} illustrates
how directional selection leads to a singular strategy at which
selection is either disruptive (left panel) or stabilizing (right
panel).

\begin{figure}
\begin{center}
\includegraphics[width=0.8\textwidth]{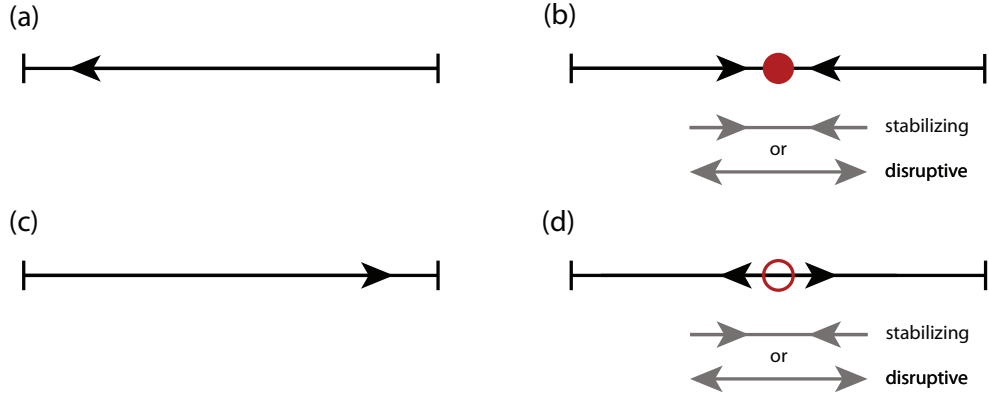}
  \end{center}
\caption{Classification of selection pressures and outcomes of gradual
  adaptation in continuous cooperation games with at most one interior
  singular strategy. If no interior singular strategy exists (panels a
  and c), cooperative investments converge to the lowest possible
  (panel a) or highest possible (panel c) levels. If an interior
  singular strategy exists (panels b and d), it may be convergence
  stable (panel b) or not (panel d). Convergence stable strategies
  (panel b) result in stationary intermediate levels of cooperation if
  selection is stabilizing (Fig.~\ref{fig1}b), or in evolutionary branching and
  thus a strategy dimorphism if selection is disruptive (Fig.~\ref{fig1}a). An
  interior singular strategy that is not convergence stable (panel d)
  separates two basins of attraction for high and low cooperative
  investments, respectively. The levels of cooperative investment
  increase along the horizontal axes, and vertical lines indicate the
  lowest possible and highest possible cooperative
  investments. Whether such limits exist is immaterial for the
  classification. Circles depict interior singular strategies where
  directional selection ceases. Filled circles represent convergence
  stable singular strategies. The dashed lines beneath panels b and d
  indicate whether selection is stabilizing or disruptive.}
\label{fig3}
\end{figure}

Assuming a univariate trait and at most one interior singular
strategy, there are only six qualitatively different configurations of
selection pressures, as shown in Figure~\ref{fig3}. In
Fig.~\ref{fig3}a and \ref{fig3}c, there is no interior singular
strategy and gradual adaptation leads to a monomorphic population of
full defectors (tragedy of the commons) or full cooperators,
respectively. In Fig.~\ref{fig3}d, the interior singular strategy is
not convergence stable (i.e., it is a repellor of monomorphic
evolution). This results in evolutionary bistability, so that the
evolutionary outcome depends on the initial condition \citep[see e.g.][]{44}. In
Fig.~\ref{fig3}b, the interior singular strategy is convergence stable
(i.e., it is an attractor of monomorphic evolution), so that gradual
adaptation leads to intermediate cooperative investments. In
Fig.~\ref{fig3}c and \ref{fig3}d, selection near the interior singular strategy
can be either stabilizing or disruptive (i.e., the interior singular
strategy is either an attractor or a repellor of dimorphic evolution,
respectively). If selection is stabilizing, it results in a locally
evolutionarily stable strategy. If selection is disruptive, the
population can become dimorphic (tragedy of the commune).

\section{Evolutionary consequences of fluctuating group size} \label{sec3}
The initial growth rate of a rare mutant player with strategy $m$ in an 
environment dominated by players with strategy $r$, as given in equation 
(\ref{eq2}), allows us to study the long-term consequences of small mutations and 
natural selection \citep{34, 8, 14}. In what follows, we first study how 
fluctuations in group size affect the location of the singular strategy. We 
show that when the payoff function does not have an explicit dependence on 
group size, the location of the singular strategy is invariant under 
fluctuations in group size. For two important classes of such payoff 
functions, with additively or multiplicatively separable benefits and costs, 
we investigate how fluctuations in group size affect the potential for 
evolutionary branching and thus for the tragedy of the commune.

\subsection{Consequences for cooperative investments}
The selection gradient $g(r)=\partial f/\partial m\vert_{m=r}$ is defined 
as the derivative of the invasion fitness $f(r,m)$ with respect to $m$ 
evaluated at $m=r$. Accordingly, the function $g(r)$ contains information 
about which nearby strategies can invade a monomorphic population of players 
with cooperative investment $r$. When the selection gradient is positive 
(negative) more (less) cooperative strategies can invade. An invading 
strategy generically replaces the resident strategy, so that the population 
again becomes monomorphic \citep{13, 15}.

From equation (\ref{eq2}), we derive the following expression for the selection 
gradient,
\[
g(r)=\left\langle k \right\rangle^{-1} \sum_{k=1}^\infty k q_k g_k 
(r),
\]
where $P_1$ and $P_2$, respectively, again denote the partial derivatives 
of $P$ with respect to its first or second argument, $\left\langle k 
\right\rangle $ is the average number of players in a group, $q_k $ is the 
probability that a group with $k$ players is formed, and $g_k (r)$ is the 
selection gradient for a fixed group size $k$,
\[
g_k (r)=P_1 (r,r,k)+\frac{1}{k}P_2 (r,r,k).
\]
More details on the derivation of the selection pressure are provided in 
Appendix B.

We now introduce $\varphi _k (r)=kg_k (r)$, so that we can apply Jensen's 
inequality, according to which the average of a convex function evaluated at 
arbitrary arguments is always larger or equal to that function evaluated at 
the average argument. Thus, if $\varphi _k (r)$ is convex (accelerating) as 
a function of $k$, then
\[
g(r)=\left\langle k \right\rangle^{-1} \sum_{k=1}^\infty \varphi_k (r)
\ge \left\langle k \right\rangle^{-1} \varphi_{\left\langle k 
\right\rangle}(r)=g_{\left\langle k \right\rangle} (r).
\]
We can thus see that fluctuations in group size imply a greater selection 
gradient. This means that cooperation will be established more rapidly and 
reach higher levels when the group size $k$ of interacting players is 
variable around $\left\langle k \right\rangle $, than when games are played 
in groups with a fixed size of $\left\langle k \right\rangle $ players. If 
$\varphi _k (r)$ is concave (decelerating) as a function of $k$, the 
opposite is true: fluctuations in player numbers then reduce the speed of 
evolutionary adjustments in cooperative investments.

Since, as we have now seen, fluctuations in group size typically affect the 
selection pressures on cooperative investments, such fluctuations can also 
shift the location of interior singular strategies. If fluctuations in group 
size shift singular strategies below the natural limit of zero cooperative 
investments, they prevent the evolution of cooperation altogether. Moreover, 
if fluctuations in group size shift singular strategies beyond the lowest 
possible or highest possible cooperative investments, they can prevent 
evolutionary bistability (if the singular strategy is not convergence 
stable) or evolutionary branching (if the singular strategy is convergence 
stable, but not locally evolutionarily stable). However, in the following we 
show that when payoffs depend only on a player's own cooperative investment 
and on its share of the group's total cooperative investment, selection 
gradients are independent of the degree of fluctuations in group size around 
a given average number of players.

\subsection{Consequences for the tragedy of the commune}
We now study the effects of fluctuating group size for games in which 
payoffs do not explicitly depend on group size,
\begin{equation}
 \label{eq3}
P\left( {r_i ,\frac{r_1 +\ldots +r_k }{k},k} \right)=P\left( {r_i ,\frac{r_1 
+\ldots +r_k }{k}} \right).
\end{equation}
For such games, the selection gradient vanishes for a singular strategy 
$r^\ast $ whenever
\begin{equation}
\label{eq4}
\left\langle k \right\rangle P_1^\ast =-P_2^\ast ,
\end{equation}
where $P_i^\ast =P_i (r^\ast ,r^\ast )$ is the partial derivative of the 
payoff function $P$ with respect to its $i$th argument, evaluated at 
$r=m=r^\ast $. Since the only feature of the distribution $q_k $ that 
appears in equation (\ref{eq4}) is the average number of players, fluctuating group 
size does not have any effect on the location of the singular strategies.

To understand the evolutionary dynamics of a monomorphic populations with a 
strategy close to the singular strategy $r^\ast $, we need to know whether 
$r^\ast $ is convergence stable (directional selection drives monomorphic 
populations toward $r^\ast )$ and whether it is locally evolutionarily 
stable (selection at $r^\ast $ is stabilizing). Near a singular strategy 
that is convergence stable but not locally evolutionarily stable, a 
monomorphic population experiences disruptive selection and will eventually 
become dimorphic through evolutionary branching. In Appendix B, we show that 
the singular strategy $r^\ast $ is convergence stable if
\begin{equation}
\label{eq5}
\left\langle k \right\rangle P_{11}^\ast +(1+\left\langle k \right\rangle 
)P_{12}^\ast +P_{22}^\ast <0,
\end{equation}
where $P_{ij}^\ast $ is the second partial derivative of the payoff function 
$P$ with respect to its $i$th and $j$th argument, evaluated at $r=m=r^\ast 
$. The singular strategy is not locally evolutionarily stable if
\begin{equation}
\label{eq6}
\left\langle k \right\rangle P_{11}^\ast +2P_{12}^\ast +\left\langle 
{k^{-1}} \right\rangle P_{22}^\ast >0,
\end{equation}
where
\[
\left\langle k^{-1} \right\rangle =\sum_{k=1}^\infty \frac{q_k}{k}.
\]
The average inverse group size $\left\langle {k^{-1}} \right\rangle $ serves 
as a measure of the strength of fluctuations in group size. It ranges from a 
minimum of $1/\left\langle k \right\rangle $ when a group of players always 
has the same size $\left\langle k \right\rangle $, to an asymptotic maximum 
of $1$ as group size becomes more and more variable.

\begin{figure}
\begin{tabular}{llll}
\raisebox{4.5cm}{(a)}
\includegraphics[width=0.3\textwidth]{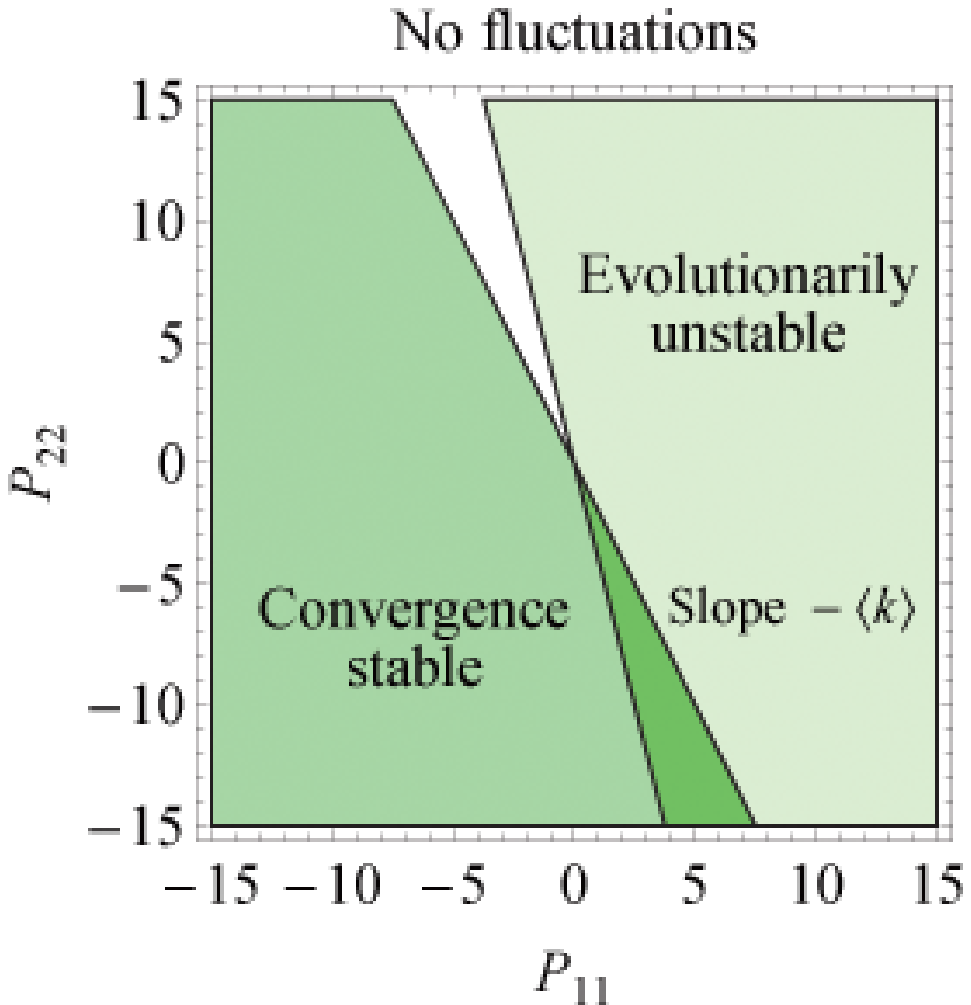} &
\includegraphics[width=0.3\textwidth]{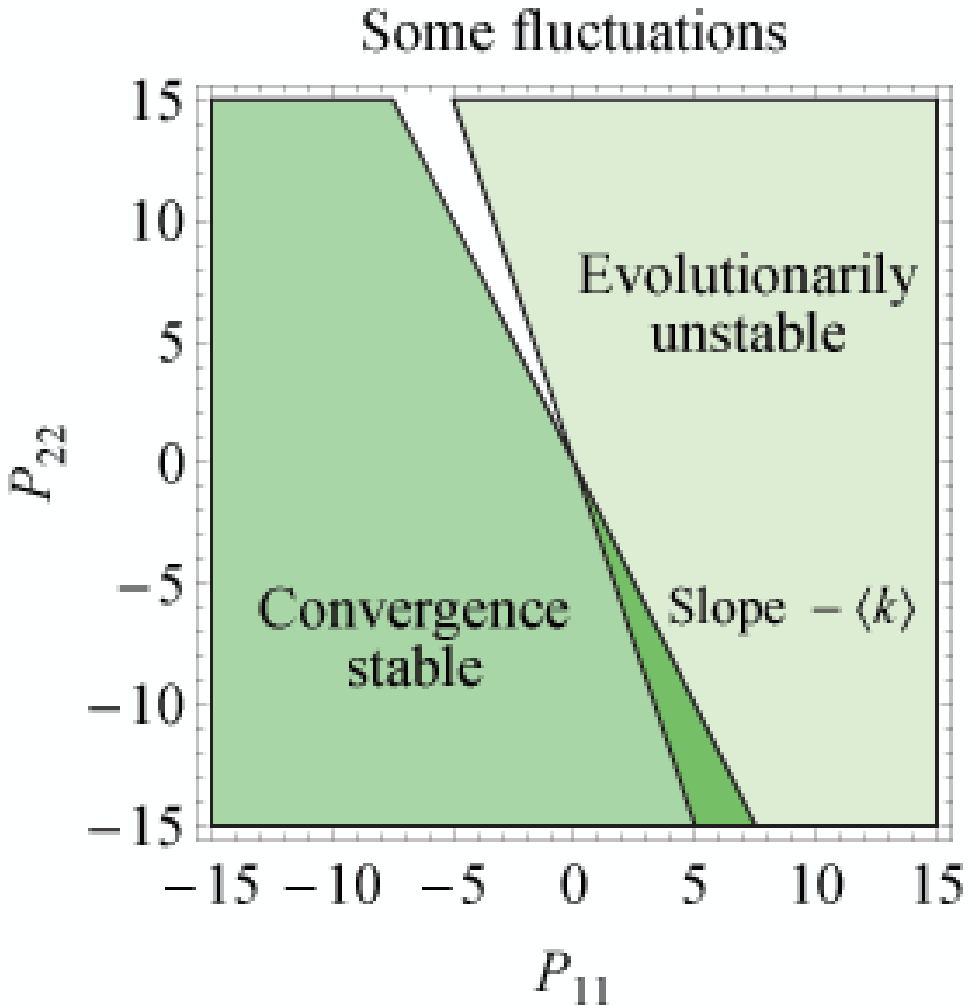} &
\includegraphics[width=0.3\textwidth]{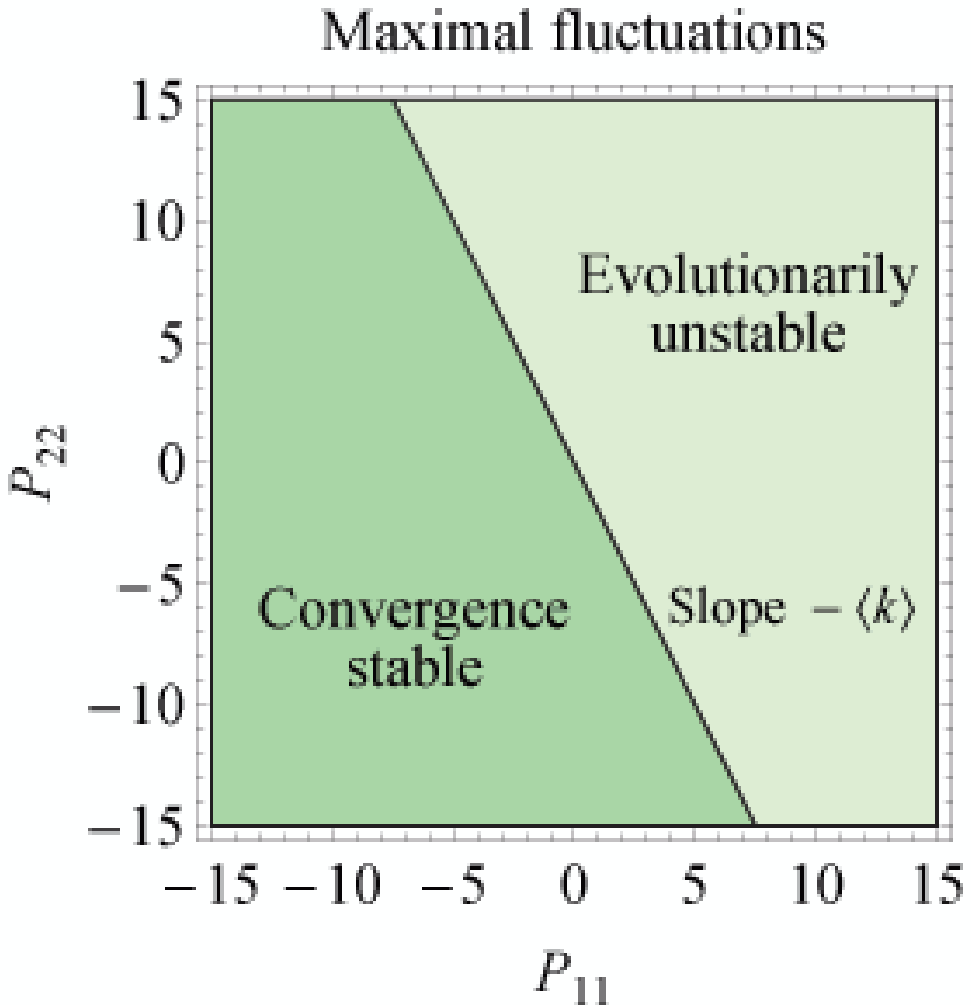}
\\ \raisebox{4.5cm}{(b)}
\includegraphics[width=0.3\textwidth]{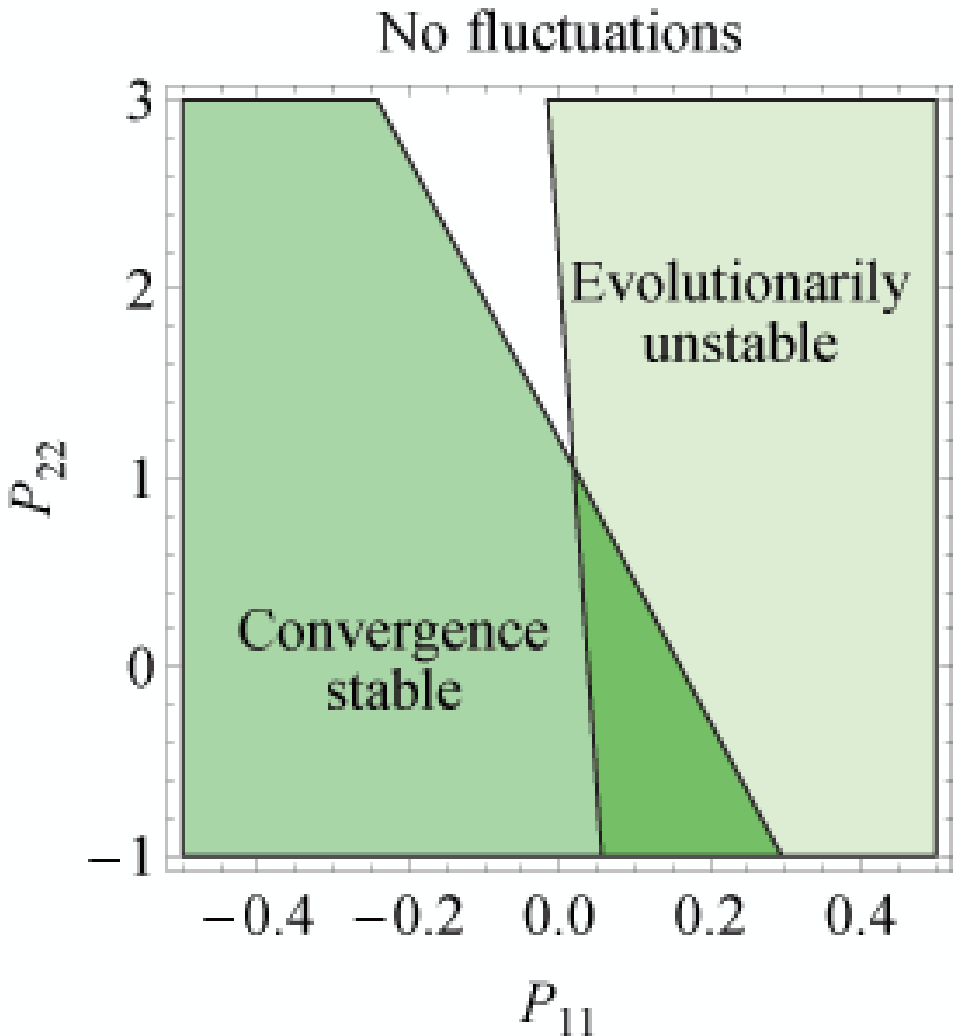} &
\includegraphics[width=0.3\textwidth]{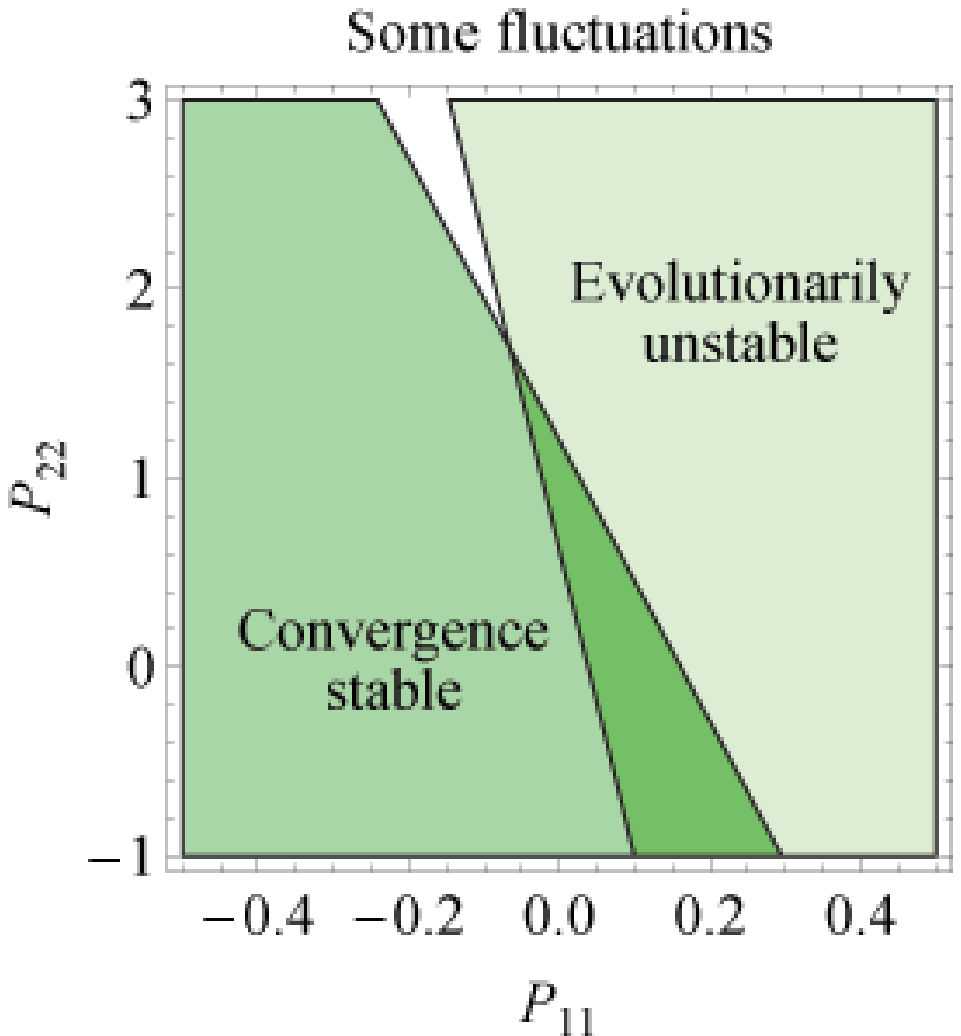} &
\includegraphics[width=0.3\textwidth]{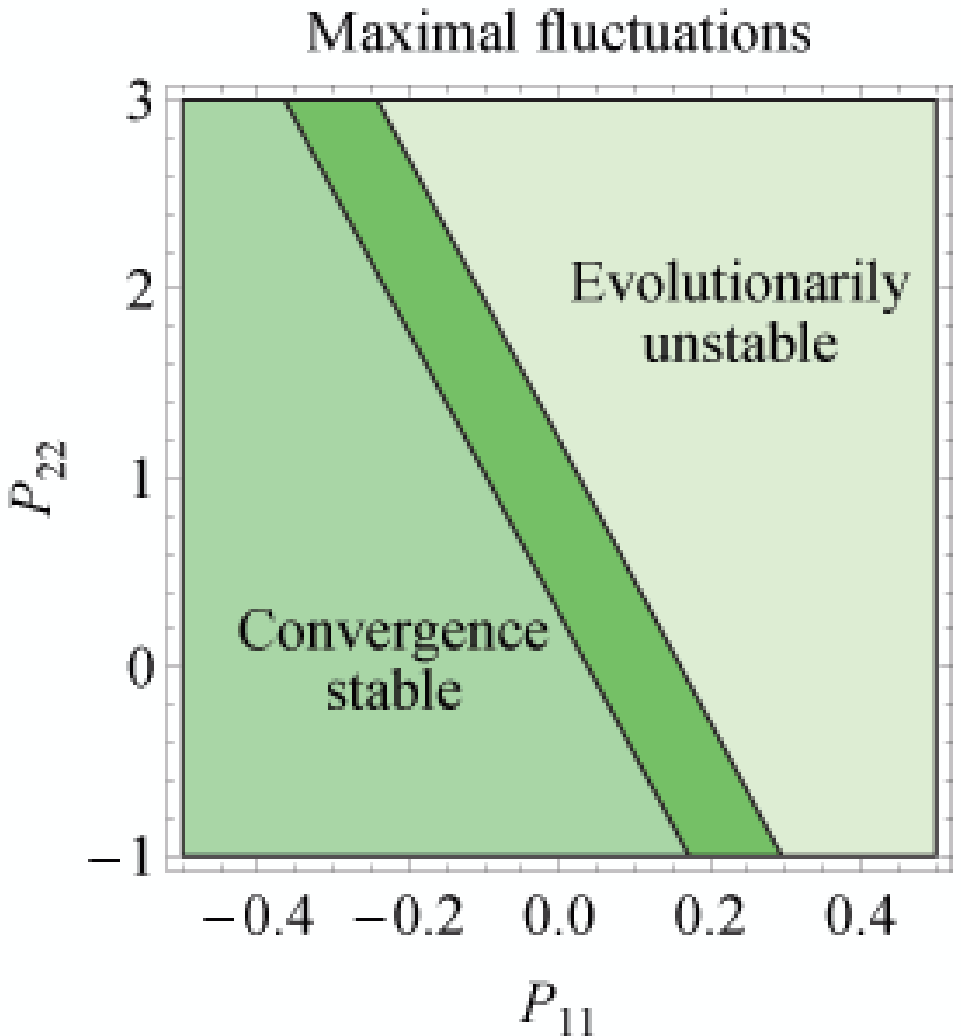}
\\ \multicolumn{3}{l}{\hfill\includegraphics[width=0.9\textwidth]{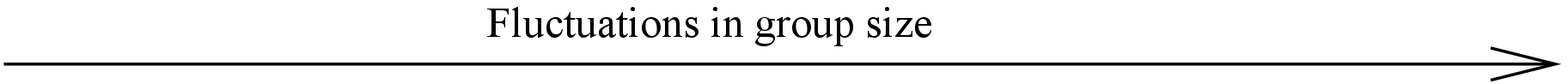}}
\end{tabular}
\caption{ Evolutionary dynamics near the singular strategy $r^*$ for
  gradually varying combinations of $P_{11}^*$ and $P_{22}^*$
  (denoting the second partial derivatives of the payoff function
  evaluated at the singular strategy) and a fixed number of players
  (left panels) or a fluctuating number of players (centre and right
  panels). The dark green regions indicate the combinations
  $(P_{11}^*, P_{22}^*)$ for which the singular strategy is
  convergence stable, inequality~(\ref{eq5}), but not locally evolutionarily
  stable, inequality~(\ref{eq6}). These are the combinations for which
  evolutionary branching eventually occurs. (a) Multiplayer extension
  of the game considered by \citet{9}. The
  benefits and costs can be separately additively and sufficient
  fluctuations in group size prevent evolutionary branching (centre
  and right panel). The regions and the cross (located at
  $P_{11}^*=3.2$ and $P_{22}^*=-11.2$) correspond to
  Fig.~\ref{fig1}b. Parameters: $P_{12}^*=0$; $\langle k \rangle=2$; $\langle k^{-1} \rangle=1/2,
  3/4, 1$ (left to right). (b) \emph{Dictyostelium} model considered
  by \citet{3}. The benefits and costs can be
  separated multiplicatively and evolutionary branching does not occur
  without fluctuating group size. The regions and the cross (located
  at $P_{11}^*=0$ and $P_{22}^*=0.79$) correspond to Fig. 4 of
  \citet{3}. Parameters: $P_{12}^*=-0.14$; $\langle k \rangle=7.5$; 
  $\langle k^{-1} \rangle= 0.13, 0.46, 1$ 
  (left to right). While the panels in (a) and (b) are produced for
  specific parameter values, the shown qualitative patterns apply to
  all games with additively and multiplicatively separable payoff
  functions, respectively. The parameters used in (b) to illustrate
  the \emph{Dictyostelium} model are not identical with those used by
  \citet{3}, but instead define a formally
  equivalent game.
}
\label{fig4}
\end{figure}

\subsubsection{Additively separable benefits and costs} \label{sec3.2.1}
We now analyze the special case in which the effects of the two arguments in 
the payoff function can be separated additively,
\begin{equation}
\label{eq7}
P\left( {r_i ,\frac{r_1 +\ldots +r_k }{k}} \right)=B\left( {\frac{r_1 
+\ldots +r_k }{k}} \right)-C(r_i ).
\end{equation}
In cooperative games, $B$ and $C$ can be interpreted as the benefit
and cost of a cooperative investment, respectively. The conditions in
inequalities~(\ref{eq1}) simply imply that both $B$ and $C$ are
increasing functions.

The separability of arguments implies that $P_{12}^\ast =0$ and it follows 
from inequality~(\ref{eq5}) that combinations of $P_{11}^\ast $ and $P_{22}^\ast $ 
for which the singular strategy is convergence stable are situated below the 
line
\[
P_{22}^\ast =-\left\langle k \right\rangle P_{11}^\ast ,
\]
which is unaffected by variation in the number of players. Likewise, it 
follows from inequality~(\ref{eq6}) that the combinations for which the singular 
strategy is not locally evolutionarily stable are situated above the line
\[
P_{22}^\ast =-\frac{\left\langle k \right\rangle }{\left\langle {k^{-1}} 
\right\rangle }P_{11}^\ast .
\]
The slope of the latter line ranges from $-\left\langle k \right\rangle ^2$ 
to $-\left\langle k \right\rangle $ as fluctuations in group size increase. 
This is shown in the three panels of Fig.~\ref{fig4}a.

The values $P_{11}^\ast $ and $P_{22}^\ast $ depend on the singular strategy 
$r^\ast $. Evolutionarily branching, and hence the tragedy of the commune, 
eventually occurs when the singular strategy is convergence stable, but not 
locally evolutionarily stable. In Fig.~\ref{fig4}a, this corresponds to the 
wedge-shaped dark green region. Since this region lies exclusively in the 
fourth quadrant, where $P_{11}^\ast =-{C}''(r^\ast )>0$ and $P_{22}^\ast 
={B}''(r^\ast )<0$, it follows immediately that the tragedy of the commune 
can only occur when both the benefit $B$ and the cost $C$ are concave around 
the singular point. Furthermore, we see that for additively separately 
payoffs fluctuating group size always reduces the parameter range in which 
evolutionary branching -- and hence the tragedy of the commune -- occurs. As 
the fluctuations increase, any point in the plane, including the cross in 
Fig.~\ref{fig4}a that corresponds to the game studied by \citet{10} (see also Fig.~\ref{fig1}), 
eventually falls outside the region in which evolutionary branching occurs. 
Hence, when payoffs are additively separable, the tragedy of the commune can 
always be avoided through sufficient fluctuations in group size.

In summary, we have shown that in games with additively separable benefits 
and costs, or more generally in games with $P_{12}^\ast =0$, fluctuations in 
group size generally reduce the scope for the tragedy of the commune to 
occur. Moreover, sufficiently large fluctuations in group size can always 
turn an evolutionary branching point into a locally evolutionarily stable 
strategy.

\subsubsection{Multiplicatively separable benefits and costs}
For games in which benefits and costs are not additively separable,
the situation can be considerably different, as illustrated by the
\textit{Dictyostelium} model studied by \citet{3}. In the
\textit{Dictyostelium} model, benefits and costs are multiplicatively
separable,
\begin{equation}
\label{eq8}
P\left( {r_i ,\frac{r_1 +\ldots +r_k }{k}} \right)=B\left( {\frac{r_1 
+\ldots +r_k }{k}} \right)C(r_i ).
\end{equation}
Using an exponentially increasing function for the benefit and a linearly 
decreasing function for the cost, \citet{3} derived an analytical condition 
demonstrating that evolutionary branching only occurs with fluctuating group 
size.

Fig.~\ref{fig4}b shows why fluctuating group size is required for evolutionary 
branching. The slopes of the lines are the same as for games with additively 
separable payoffs, but the line with constant slope now intercepts the 
$P_{22}^\ast$-axis at the point $-(1+\left\langle k \right\rangle 
)P_{12}^\ast$, while the remaining line intercepts at $-2P_{12}^\ast 
/\left\langle {k^{-1}} \right\rangle $. The effect of fluctuating group size 
thus depends on the sign of $P_{12}^\ast $. For multiplicatively separable 
payoffs, $P_{12}^\ast $ is always negative, $P_{12}^\ast ={B}'(r^\ast 
){C}'(r^\ast )<0$. Without fluctuations in group size, the region in which 
evolutionary branching occurs lies entirely in the fourth quadrant. As 
fluctuations increase, the intercept $-2P_{12}^\ast /\left\langle {k^{-1}} 
\right\rangle $ decreases from $-\left\langle k \right\rangle ^2P_{12}^\ast 
$ to $-\left\langle k \right\rangle P_{12}^\ast$, which is below 
$-(1+\left\langle k \right\rangle )P_{12}^\ast$. Thus, \citet{3} did not find 
evolutionary branching without fluctuations in group size, because they used 
a linear cost function, which implies $P_{11}^\ast =0$. For sufficiently 
small values of $P_{22}^\ast $, evolutionary branching cannot occur without 
fluctuating group size.

The case $P_{12}^\ast >0$ is similar to games with additively separable 
benefits and costs. In particular, sufficiently large fluctuations in group 
size can always turn an evolutionarily branching point into a locally 
evolutionarily stable strategy, hence avoiding the tragedy of the commune.

\subsection{General classification of the consequences of fluctuating group size}
The preceding analysis shows that the sign of $P_{12}^\ast $ has a profound 
impact on the evolutionary dynamics. For $P_{12}^\ast \ge 0$, fluctuations 
in the number of players always reduces the scope for evolutionary 
branching, whereas for $P_{12}^\ast <0$, other outcomes are possible.

To understand the effects of fluctuating group size in general cooperation 
games with payoffs that are neither additively nor multiplicatively 
separable, we combine inequalities~(\ref{eq5}) and (\ref{eq6}), resulting in a condition for 
the singular strategy to be convergence stable but not locally 
evolutionarily stable,
\begin{equation}
\label{eq9}
\left\langle k \right\rangle P_{12}^\ast +P_{22}^\ast <-\left\langle k 
\right\rangle P_{11}^\ast -P_{12}^\ast <P_{12}^\ast +\left\langle {k^{-1}} 
\right\rangle P_{22}^\ast .
\end{equation}
In this double inequality, only the rightmost term changes with the 
variability in group size. This allows us to obtain a necessary condition 
for evolutionary branching when the number of players is fixed,
\begin{equation}
\label{eq10}
\left\langle k \right\rangle P_{12}^\ast +P_{22}^\ast <P_{12}^\ast 
+\frac{1}{\left\langle k \right\rangle }P_{22}^\ast .
\end{equation}
Writing $A=\left\langle k \right\rangle P_{12}^\ast +P_{22}^\ast $ and 
$B=P_{12}^\ast +\left\langle k \right\rangle ^{-1}P_{22}^\ast $ for the 
left-hand and right-hand side, respectively, we can classify a game 
according to whether evolutionary branching is possible without variation in 
group size ($A<B)$ or not ($A>B)$. Analogously, we obtain a necessary 
condition for evolutionary branching when fluctuations in the number of 
players are maximal,
\[
\left\langle k \right\rangle P_{12}^\ast +P_{22}^\ast <P_{12}^\ast 
+P_{22}^\ast ,
\]
which simplifies to $P_{12}^\ast <0$. Thus, we may further classify a game 
according to whether evolutionary branching is possible with maximal 
fluctuation in group size ($P_{12}^\ast <0)$ or not ($P_{12}^\ast <0)$.

\begin{figure}
\centering
\includegraphics[width=0.8\textwidth]{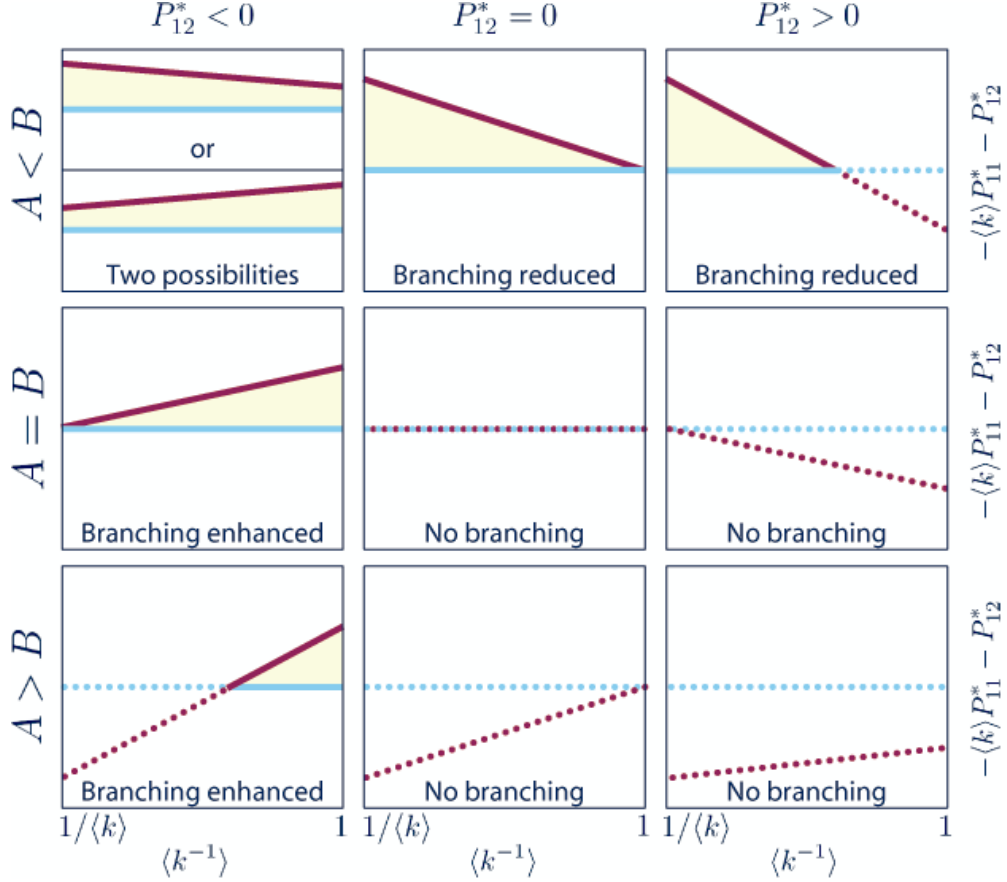}
\caption{The effects of fluctuating group size on the occurrence of
  evolutionary branching in general cooperation games with payoff
  functions not explicitly depending on the group size. Qualitatively
  distinct cases can be classified according to the values of
  $P_{12}^*$, $A = \langle k \rangle P_{12}^* + P_{22}^*$ and $B =
  P_{12}^* + \langle k \rangle^{-1} P_{22}^*$. Each panel shows, as a
  function of $\langle k^{-1} \rangle$, the range of values in which the middle part
  of inequality (\ref{eq9}), $\langle k \rangle P_{11}^*-P_{22}^*$, must lie
  for evolutionary branching to occur. The blue (medium gray) and red
  (dark gray) lines represent the left-hand and right-hand sides of
  inequality~(\ref{eq9}) respectively. When inequality~(\ref{eq9}) can be fulfilled,
  these lines are shown as continuous, otherwise as dotted. The
  regions in which inequality~(\ref{eq9}) are fulfilled are depicted in yellow
  (light grey). For $P_{12}^*\geq 0$, fluctuations in group size
  always reduces the scope for evolutionary branching. If instead
  $P_{12}^* < 0$, the effect of fluctuating group size depends on
  whether $A \leq B$ or $A > B$. In the latter case, the region in
  which evolutionary branching occurs increases with variation in the
  number of players, whereas in the former case information about
  $P_{22}^*$ is required to determine the effects of fluctuating group
  size. }
\label{fig5}
\end{figure}

Fig.~\ref{fig5} provides a graphical representation of these classifications. In each 
case, the range of values in which the middle point of inequality~(\ref{eq9}) must 
lie for evolutionary branching to occur is shown as a function of the 
variability in group size. For $P_{12}^\ast <0$, and thus for games in which 
the benefits and costs are multiplicatively separable, the effects of 
fluctuating group size depend on whether inequality~(\ref{eq10}) holds. If it does 
not hold, fluctuating group size always reduces the scope for evolutionary 
branching. If it does hold, fluctuating group size may prevent or induce 
evolutionary branching. To distinguish between these two cases, we need to 
check whether
\[
P_{12}^\ast +\left\langle {k^{-1}} \right\rangle P_{22}^\ast <P_{12}^\ast 
+P_{22}^\ast ,
\]
which applies if $P_{22}^\ast >0$. In this case, fluctuations in group size 
hinder evolutionary branching. If instead $P_{22}^\ast <0$, fluctuations in 
group size promote evolutionary branching.

For games in which benefits and costs can be separated additively, 
$P_{12}^\ast =0$, and we infer from Fig.~\ref{fig5} that fluctuations in the number 
of players always reduces the scope for evolutionary branching, in line with 
the conclusions in Sect.\ref{sec3.2.1}.

\section{Conclusions}
Fluctuations are an inherent feature of social and natural systems. We have 
shown that fluctuating group size can have important consequences for the 
evolution of cooperation, by impacting both the level and the evolutionary 
stability of cooperative investments. The former impact can turn stable 
intermediate cooperative investments into a tragedy of the commons, by 
shifting the predicted cooperative investments below the natural lower limit 
of no investment. The latter impact can turn stable intermediate cooperative 
investments into the recently elucidated tragedy of the commune. Naturally, 
the consequences of fluctuating group size may also work in the opposite 
direction, preventing the tragedy of the commons or the tragedy of the 
commune.

Our results in Sect.~\ref{sec3} on how convergence stable cooperative
investments are expected to shift with fluctuations in group size are
based on the assumption that a player's payoff may be affected by its
own investment, the share it receives of the group's total investment,
and the group's size. We have shown that in games with payoff
functions that do not explicitly depend on group size, convergence
stable cooperative investments do not shift with fluctuations in group
size. For this class of games, the consequences of fluctuations in
group size are thus limited to altering the evolutionary stability of
cooperative investments. We have also shown that these consequences
for potential evolutionary branching, and hence for the tragedy of the
commune, can be understood in terms of the mixed second partial
derivative $P_{12}^\ast $ of the payoff function. To obtain
$P_{12}^\ast $, the payoff function $P$ is differentiated with respect
to a player's investment and with respect to the share it receives of
the group's total investment, with the resultant derivative being
evaluated at the evolutionarily singular strategy at which the
directional selection pressures on cooperative investments
cease. Fluctuations in group size make evolutionary branching less
likely if $P_{12}^\ast \ge 0$, and we have shown that sufficient
fluctuations in group size can always turn disruptive selection into
stabilizing selection, hence preventing the tragedy of the commune. If
$P_{12}^\ast <0$, fluctuations in group size can either turn
disruptive selection into stabilizing selection or vice versa, and
further information is needed to determine which of these alternative
consequences occurs. Most theoretical studies of cooperation are based
on games with $P_{12}^\ast =0$ \citep[which is guaranteed when the
payoff function is additively separable; e.g.,][]{10, 31} or on
games with $P_{12}^\ast <0$ \citep[which is guaranteed when the payoff
function is multiplicatively separable; e.g.,][]{41, 3, 12, 5, 4}.

The results obtained in this study demonstrate that fluctuations in group 
size can significantly affect cooperation evolution in real-world systems by 
increasing or decreasing, and by stabilizing or destabilizing, cooperative 
investments. Yet, to date only a handful of studies have considered the role 
of fluctuating group size for the evolution of cooperation \citep{36, 24, 23, 22}, 
so that many interesting questions as yet remain unexplored. Two extensions 
of our work here may be of particular relevance. A first promising direction 
is to consider the consequences of fluctuating group size on the tragedy of 
the commune for cooperation games with payoff functions that explicitly 
depend on group size, which includes situations in which a group's total 
investment is not divided up equally between players. A second promising 
direction is to consider processes according to which players adopt new 
strategies that differ from the gradual-adaptation process we have analyzed 
here.

Studies of cooperation games with continuous strategies have recently 
started to add greater realism and new vistas to cooperation research. The 
future promise of these developments will be further enhanced by accounting 
for the fluctuations in group size typically arising in realistic 
multiplayer games.

\section*{Acknowledgements}
We thank Duncan Greig, Karl Sigmund, and David Sumpter for valuable 
discussions and encouragement. This research has been supported by the 
European Marie Curie Research Training Network FishACE (Fisheries-induced 
Adaptive Changes in Exploited Stocks), funded through the European 
Community's Sixth Framework Programme (Contract MRTN-CT-2004-005578). U.D. 
gratefully acknowledges additional financial support by the Specific 
Targeted Research Project FinE, the European Science Foundation, the 
Austrian Science Fund, the Austrian Ministry of Science and Research, and 
the Vienna Science and Technology Fund.
 
\section*{Appendix A: Demographic dynamics}
We consider a population of players in which $n$ distinct cooperation 
strategies $r_1 ,\ldots ,r_n $ are represented with densities $X_1 ,\ldots 
,X_n $. The demographic dynamics of these densities is assumed to be given 
by
\[
\frac{\dot {X}_i }{X_i }= R(r_i ,X)-\mu (X_1 +\ldots +X_n ),
\]
where $\dot {X}_i =dX_i /dt$, $R(r_i ,X)$ is the expected number of players 
with strategy $r_i $ emerging from a game, and the loss rate $\mu $ is 
included to allow for density regulation. We scale time $t$ such that the 
rate at which a player participates in a game equals 1. Assuming that a 
player participates in a $k$-player game with probability $p_k $ and that 
the participants of such games are drawn at random from the population, 
$R(r_i ,X)$ is given by

\begin{align*}
&R(r_i, X) = \\ &\sum_{k=1}^\infty p_k \sum_{1 + k_1 + \ldots + k_n = k} 
  \frac{(k-1)!}{k_1! \ldots k_n!} \frac{X_1^{k_1}\ldots X_n^{k_n}}{(X_1 + \ldots + X_k)^{k-1}} P\left(r_i, \frac{k_1 r_1 + \ldots + (k_i+1)r_i + \ldots + k_n r_n}{k}, k\right),
\end{align*}
where $P$ is the payoff function defined in Sect. 2.2. The second sum above 
reflects the fact that the one focal player with strategy $r_i $ 
participates in a $k$-player game with $k_1 ,\ldots ,k_n $ other players 
that follow strategies $r_1 ,\ldots ,r_n $. With $x_i =X_i /(X_1 +\ldots 
+X_n )$ denoting the frequency of strategy $r_i $ in the population, we 
obtain
\begin{equation}
\label{eq11}
\frac{\dot {x}_i }{x_i }=\frac{\dot {X}_i }{X_i }-\frac{\dot {X}_1 +\ldots 
+\dot {X}_n }{X_1 +\ldots +X_n }=R(r_i ,x)-[x_1 R(r_1 ,x)+\ldots +x_n R(r_n 
,x)],
\end{equation}
which is a generalization of the classical replicator equation \citep{25}.

\section*{Appendix B: Evolutionary dynamics}
From equation (\ref{eq11}) we deduce the initial increase in the frequency of a rare 
mutant strategy $m$ in an environment dominated by players with strategy 
$r$. Writing $x_1 =x_r $, $x_2 =x_m $, $r_1 =r$, and $r_2 =m$ when only 
these two cooperation strategies are present, we have
\[
R(m, x) = \sum_{k=1}^\infty p_{k} 
\sum_{j = 0}^{k-1} \binom{k-1}{j} x_r^j x_m^{k-1-j} P\left(m, \frac{jm+(k-j)r}{k}, k\right).
\]
The invasion fitness of the rare mutant morph is then defined as
\[
f(r,m)=\mathop {\lim }\limits_{x_m \to 0+} \frac{\dot {x}_m }{x_m 
}=R(m,(1,0))-R(r,(1,0)),
\]
which gives
\begin{equation}
\label{eq12}
f(r,m)=\sum_{k=1}^\infty p_k \left[ P\left(m,\frac{m+(k-1)r}{k},k 
\right)-P\left(r,r,k \right) \right].
\end{equation}
We can alternatively express the further calculations in terms of the 
probability $q_k $ that a game involves $k$ players, which is related to the 
individual's probability $p_k $ of joining a $k$-player game by
\[
p_k =\frac{kq_k }{\sum_{k=1}^\infty k q_k} = \frac{kq_k 
}{\left\langle k \right\rangle },
\]
where $\left\langle k \right\rangle $ is the average group size, 
$\left\langle k \right\rangle =\sum\nolimits_{k=1}^\infty {kq_k } $.

From the invasion fitness $f(r,m)$ in equation (\ref{eq12}), we obtain the selection 
gradient
\[
g(r)=\left. \frac{\partial f(m,r)}{\partial m} \right|_{m=r} = \left\langle k 
\right\rangle^{-1}\sum_{k=1}^\infty q_k\left[kP_1 (r,r,k)+P_2 
(r,r,k) \right],
\]
where $P_i $ denotes the first partial derivative of $P$ with respect to its 
$i$th argument. Of particular interest are the singular strategies $r^\ast $ 
at which directional selection ceases, $g(r^\ast )=0$.

A singular strategy $r^\ast $ is convergence stable, and nearby monomorphic 
populations will thus evolve toward it, if
\begin{equation}
\label{eq13}
g'(r^\ast )=\left\langle k \right\rangle^{-1}\sum_{k=1}^\infty q_k \left[ kP_{11}^\ast (k)+(1+k)P_{12}^\ast (k)+P_{22}^\ast (k) \right]<0.
\end{equation}
Here, $P_{ij} $ denotes the second partial derivative of $P$ with respect to 
its $i$th and $j$th arguments, and the asterisks indicate that these 
derivatives are evaluated at the singular strategy, $P_{ij}^\ast (k)=P_{ij} 
(r^\ast ,r^\ast ,k)$.

A singular strategy $r^\ast $ is not locally evolutionarily stable, and 
selection will thus be disruptive in its vicinity, if
\begin{equation}
\label{eq14}
\left. \frac{\partial^2f(m,r)}{\partial m^2} \right|_{m=r=r^\ast}
=\left\langle k \right\rangle^{-1} \sum_{k=1}q_k \left[ 
kP_{11}^\ast (k)+2P_{12}^\ast (k)+k^{-1}P_{22}^\ast (k) \right]>0.
\end{equation}
We can combine inequalities (\ref{eq13}) and (\ref{eq14}) into a single criterion for the 
occurrence of an evolutionary branching point,
\begin{align*}
 \sum_{k=1}^\infty q_k \left[ (1+k)P_{12}^\ast (k)+k^{-1}P_{22}^\ast 
(k) \right] & <-\sum_{k=1}^\infty q_k P_{11}^\ast (k) \\ 
 & <\sum_{k=1}^\infty \left[ 2P_{12}^\ast (k)+k^{-1}P_{22}^\ast 
(k) \right]. 
\end{align*}
The inequality on the left is the condition for convergence stability 
(implying evolutionary attraction toward $r^\ast )$, while the inequality on 
the right is the condition for the lack of local evolutionary stability 
(implying disruptive selection at $r^\ast )$.

\end{document}